\documentclass[pre,epsfig,twocolumn,showpacs,preprintnumbers,amssymb]{revtex4}
\usepackage{bm,graphicx,amsmath}

\begin{document}

\title{Diffusive transport of light in three-dimensional disordered Voronoi structures}

\author{Zeinab Sadjadi}
\affiliation{Institute for Advanced Studies
in Basic Sciences (IASBS), P. O. Box 45195-1159, Zanjan 45195, Iran}

\author{MirFaez Miri }
\email{miri@iasbs.ac.ir} \affiliation{Institute for Advanced Studies
in Basic Sciences (IASBS), P. O. Box 45195-1159, Zanjan 45195, Iran}

\author{Holger Stark}
\affiliation{Technische Universit\"at Berlin, Institut f\"ur Theoretische
Physik, Hardenbergstr. 36, D-10623 Berlin, Germany}

\begin{abstract}
The origin of diffusive transport of light in dry foams is still
under debate. In this paper, we consider the random walks of
photons as they are reflected or transmitted by liquid films
according to the rules of ray optics. The foams are approximately
modeled by three-dimensional Voronoi tessellations with varying
degree of disorder. We study two cases: a constant intensity
reflectance and the reflectance of thin films. Especially in the
second case, we find that in the experimentally important regime
for the film thicknesses, the transport-mean-free path $l^{\ast}$
does not significantly depend on the topological and geometrical
disorder of the Voronoi foams including the periodic Kelvin foam.
This may indicate that the detailed structure of foams is not
crucial for understanding the diffusive transport of light.
Furthermore, our theoretical values for $l^{\ast}$ fall in the
same range as the experimental values observed in dry foams. One
can therefore argue that liquid films contribute substantially to
the diffusive transport of light in {dry} foams.
\end{abstract}



\maketitle

\section{Introduction}
The interaction of light with matter is a highly interesting
subject with many facets. For example, multiply scattered light in
turbid media is a very complex topic. However, it becomes
treatable when, after a sufficiently large amount of scattering
events, light reaches its diffusive limit\ \cite{sheng}. Diffusing
light is used for biomedical imaging\ \cite{Yodh95} and in
diffusing-wave spectroscopy (DWS) it is able to monitor dynamic
processes in turbid materials\ \cite{DWS}. Recently, even clear
evidence for strong localization of light was given\
\cite{Aegerter06}. In colloidal suspensions and nematic liquid
crystals\ \cite{DWSnematic} the respective scattering events and
diffusion of light are well understood. But why are aqueous foams\
\cite{riv, Weaire1999}, as used, e.g, in shaving cream, turbid?

Aqueous foams consist of gas bubbles separated by liquid films\
\cite{riv, Weaire1999}. In dry foams, where most of the liquid has
drained out, the bubbles are deformed to polyhedra. The so-called
Plateau rules state that always three films meet in the Plateau
borders, which form a network of liquid channels throughout the
foam, and that four of these borders meet at tetrahedral vertices.
Foams are visibly opaque although both the gas and liquid
components are transparent.

Precise light-scattering experiments show that light transport reaches its
diffusive limit in foams\ \cite{Durianold, durApp, rheo},
which means that photons perform a random walk. However, the mechanisms
underlying this random walk are not well understood.
One suggestion is that scattering from the Plateau borders is
responsible for the diffusing light\
\cite{durApp}.
The transport-mean-free path $l^{\ast}$, over which the photon direction
becomes randomized, is then predicted as
$ l^{\ast} \propto H/\sqrt{\varepsilon} $, where $H$ is the average bubble
diameter and $\varepsilon$ is the liquid volume fraction. However, data rule
this out in favor of the empirical law\ \cite{durApp,Gittings2004}
\begin{equation}
l^{\ast} \approx H(\frac{0.14}{\varepsilon}+1.5 ) . \label{emp}
\end{equation}
The authors of Ref.\ \cite{Gittings2004} state that this may imply
significant contributions from scattering from vertices\
\cite{Skipetrov02} or films\ \cite{miriA, miriB, miriC, miriD,
miriE}. Novel transport effects, such as total internal reflection
of photons inside the Plateau borders, are studied in\
\cite{Gittings2004, miriF}.

In recent years we have explored how liquid films in combination
with ray optics determine light transport in dry foams\
\cite{miriA,miriB, miriC, miriD, miriE}. In a step-by-step
approach, we employed three model foams: the honeycomb structure,
two-dimensional Voronoi tessellations\ \cite{okabe}, and finally
the three-dimensional Kelvin structure\ \cite{Weaire1999,
Weaire1996}. Apparently, the {\it periodic} Kelvin structure is
highly idealistic. In his momentous experimental study of bubble
shapes, Matzke did not find even a single Kelvin cell\
\cite{matz}. Therefore, in this paper, we extend our studies
towards real dry foams. We introduce {\it topological and
geometrical disorder} based on a three-dimensional Voronoi foam
model to investigate the influence of disorder. Note that a
Voronoi foam does not obey Plateau's rules, and indeed all films
of a Voronoi foam are flat. Thus an accurate sample foam provided
by the {\it Surface Evolver} software\ \cite{su} clearly better
describes the geometry of a real foam. Nevertheless, we
deliberately concentrate on the Voronoi tessellations with flat
films in order to greatly simplify our simulations of light
transport, which are still very time consuming. Moreover, a
comparison of the results presented in this paper with future
theoretical studies of diffusive light transport in exact models
of real foams, will highlight the role of both their geometry and
their curved films.

\begin{table*}[t]
\caption{\label{t1} Topological and geometrical characteristics of
the cells in the Kelvin foam, our Voronoi foams 1-5, real foams\
\cite{mon}, simulated foams\ \cite{kray1}, and the Poisson Voronoi
tessellation (PVT)\ \cite{okabe}. The symbols are explained in the
text.}
\begin{center}
\begin{tabular}{c c c c c c c c c c}\hline \hline
 & Kelvin Foam & Foam 1 & Foam 2    &Foam 3 & Foam 4 & Foam 5 & Real Foam & Simulation & PVT\\ \hline
$\Delta L/L$           & 0     & 0.050197  & 0.126543  &0.164922   &0.312005   &0.486383   &-  &- &-\\
$\Delta S/S$           & 0     & 0.00684   &0.01768    &0.02340    &0.04330    &0.07433    &-  &-  &-\\
 $\Delta V/V$           & 0     & 0.011179  & 0.028264  &0.037067   &0.069619   &0.122512   &0.08-0.29  &- &-\\
 $I_{Q}$            & 0.757 & 0.752663  &0.748894   &0.745840   &0.727823   &0.686616   &0.694      &0.693-0.751&-\\
 $ \langle f \rangle$ &14   &14.0       & 14.0      &14.0       & 14.007812 &14.188232  &13.5       &13.7-13.94&15.535\\
 $ \langle n \rangle$ & 5.142857    &5.142857 &5.142857  &5.142857  &5.143335   &5.154229   &5.11   &-&5.228\\
 $\mu_{2,f}$        & 0     & 0.0       &0.0        &0.0        & 0.028259  &0.930633   &0.85-3.66  &0.812-1.46&11.055\\
 $\mu_{2,n}$        & 0.979593     & 0.979593  &0.979593   &0.979593   & 0.981509  &1.028764   &0.365-0.42     &-&2.49\\
 \hline \hline
\end{tabular}
\end{center}
\end{table*}

Cells in a foam are much larger than the wavelength of light, thus
one can employ ray optics and follow a light beam or photon as it
is reflected by the liquid films with a probability $r$ called the
intensity reflectance. We perform  extensive simulations to
determine the diffusive limit of light for two models. In model I,
we choose a constant intensity reflectance $r$ to explore the
effect of disorder. The essential result of our Monte-Carlo
simulations is summarized in the empirical formula for the
transport-mean-free path $l^{\ast}$,
\begin{equation}
l^{\ast}_{\text{Voronoi}}(r)= 0.63 {H} \frac{ 1-r}{r}   (1-b_{1} +
b_{2} r) .
\label{diffVoronoi}
\end{equation}
The main behavior of $l^{\ast}_{\text{Voronoi}}(r)$ is governed by
the factor $(1-r)/r$. Quite remarkably, this factor is also found
in the honeycomb, two-dimensional disordered Voronoi, and
three-dimensional regular Kelvin structures. However, there is a
small but systematic deviation from $(1-r)/r$, described by the
last factor in Eq.\ (\ref{diffVoronoi}) with $0<b_{1}<0.1$ and
$b_{2} \approx 0.1$. Both constants show a slight dependence on
disorder in the Voronoi foam. In model II, we use the intensity
reflectance of thin films with its significant dependence on the
incident angle and on $d/\lambda$, where $d$ is film thickness and
$\lambda$ the wavelength of light.
Our theoretical values for the transport-mean-free path $l^{\ast}$
fall in the same range as the experimental values for the driest
foams in Ref.\ \cite{durApp}. This shows that liquid films are
important for the understanding of photon diffusion in dry foams
otherwise we should have obtained much larger transport-mean-free
paths. We also observe that topological and geometrical disorder
do not change the qualitative behavior of $l^{\ast}$. Even
quantitative changes are not very pronounced. Thus, our extensive
numerical simulations support a conclusion that we already drew
from our previous two dimensional investigations\ \cite{miriB}:
the detailed structure of a Voronoi foam is {\it not} important
for understanding the diffusive limit of light transport.

Our article is organized as follows. In Section\ \ref{model} we
introduce the Voronoi tessellation as a simple model foam. Photon
transport in a Voronoi structure using constant and thin-film
intensity reflectances are discussed in Secs.\ \ref{rconst} and\
\ref{FreFre}, respectively. Discussions, conclusions, and an
outlook are presented in Sec.\ \ref{dis}.

\section{Model foam} \label{model}
As a {\it simple} model for a three-dimensional disordered dry
foam, we choose the Voronoi tessellation\ \cite{okabe}. Voronoi
foams satisfy the topological requirements on edge and face
connectivity in Plateau's rules, but not the geometric conditions;
e.g., the angles between two edges are not equal to the
tetrahedral angle. To make such a Voronoi foam, a distribution of
seed points in the simulation box is chosen and then Voronoi
polyhedrons are constructed in complete analogy to the
Wigner-Seitz cells for periodic lattice sites. We start with a
body-centered cubic lattice of seed points, which gives the Kelvin
foam, and then systematically introduce disorder by shifting the
seed points in random directions. The magnitude of the
displacement vectors of the seed points is uniformly distributed
in the interval $[0, h]$ and disorder in a Voronoi foam increases
with $h$. Referring $h$ to the diameter $H_{\text{Kelvin}}$ of a
cell in the Kelvin foam\ \cite{miriD}, we study five samples with
$h/H_{\text {Kelvin}} = 0.02, 0.05,0.07,0.12,$ and 0.20. Our
Voronoi foams are produced by the software {\it Qhull}\
\cite{Qhull}. Typically, the simulation box contains around $8200$
cells, $62800$ films, and $11200$ edges. To simulate the diffusion
of photons in these model foams, periodic boundary conditions are
implemented.

In disordered foams several random variables exist; e.g., the edge
length $L$, the cell surface $S$, the cell volume $V$, the number
of faces per cell $f$, and the number of edges per face $n$. A
first insight into the cellular structure can be gained through
the {\it distribution} of these variables. To characterize foams
and their geometrical and topological disorder, we have collected
data for the following quantities:
\begin{eqnarray}
\Delta L/{L} &=& \frac{[\langle L^2 \rangle-{\langle  L
\rangle}^2]^{1/2}}{ \langle L \rangle  } ,\nonumber \\
\Delta S/{S} &=& \frac{[\langle S^2 \rangle-{\langle  S
\rangle}^2]^{1/2}}{ \langle S \rangle  } ,\nonumber \\
\Delta V/{V} &=& \frac{[\langle V^2 \rangle-{\langle  V
\rangle}^2]^{1/2}}{ \langle V \rangle  } ,\nonumber \\
 I_{Q} &=& \langle \frac{ 36 \pi V^2}{  S^3 }\rangle \nonumber \\
 \mu_{2,f} &=&\langle f^2 \rangle-{\langle f \rangle}^2  ,\nonumber \\
\mu_{2,n} &=& \langle n^2 \rangle-{\langle n \rangle}^2,
\end{eqnarray}
where averages are denoted by $\langle~\rangle$. The first three
quantities give the standard deviations relative to mean values of
the three respective distributions for $L$, $S$, and $V$. The
isoperimetric quotient $I_{Q}$ is one for a sphere and, therefore,
measures how strongly the cells are deformed relative to a sphere.
Finally, the variances $\mu_{2,f}$ and $\mu_{2,n}$ are considered
as measures for topological disorder. We also note that instead of
$ \Delta V/{V} $ and $ I_{Q}$, alternative measures $p = { \langle
{ V} \rangle^{2/3} }/{\langle {V}^{2/3} \rangle} -1 $ and
$\beta=\langle S/(36 \pi V^2)^{1/3} \rangle $, are introduced in
Ref.\ \cite{krayPRL}.

It is instructive to compare our samples with slightly
polydisperse real foams investigated by Monnereau {\it et al.}\
\cite{mon}, and random monodisperse foams simulated by Kraynik
{\it et al.}\ \cite{kray1}, see Table\ \ref{t1}. To produce such
foams, Kraynik {\it et al.} used the Surface Evolver to relax an
initial Voronoi structure produced from the packing of spheres.
Another structure that is commonly considered the complete spatial
random pattern, is the Poisson Voronoi tessellation (PVT)\
\cite{okabe}. It allows the analytic evaluation of the moments of
various distribution functions, and consequently has gained much
attention. However, PVT clearly cannot serve as a model for real
foams. As Table\ \ref{t1} shows, PVT possesses a high degree of
topological disorder, $\mu_{2,f} = 11.055$ and $\mu_{2,n}=2.49$,
while in real foams $\mu_{2,f}<3.66$ and $\mu_{2,n}<0.42$\
\cite{mon}. Moreover, in contrast to the simulated foams in\
\cite{kray1}, PVT has an overwhelming number of short edges, i.e.,
the distribution of edge lengths reaches its maximum at $L=0$\
\cite{okabe}.

\begin{figure}
\includegraphics[width=0.8\columnwidth]{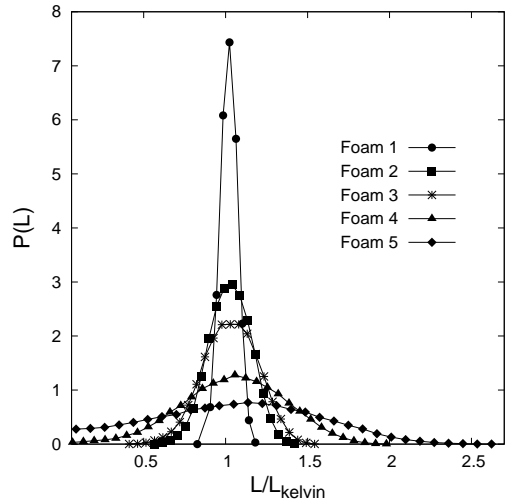}
\caption{Distribution of edge length (in units of
$L_{\text{Kelvin}}$) for the disordered Voronoi foams 1-5.}
\label{v1}
\end{figure}

\begin{figure}
\includegraphics[width=\columnwidth]{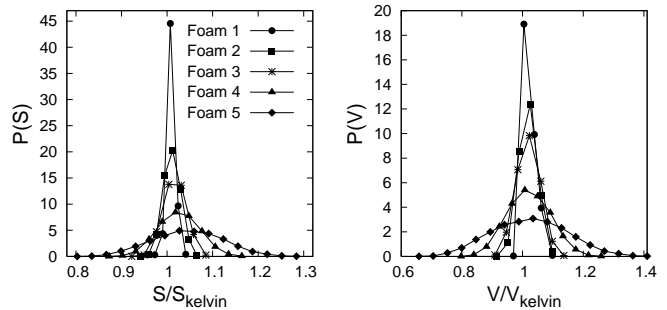}
\caption{Distribution of cell surface (in units of
$S_{\text{Kelvin}}$), and distribution of cell volume (in units of
$V_{\text{Kelvin}}$), for the disordered Voronoi foams 1-5.}
\label{v2}
\end{figure}

\begin{figure}
\includegraphics[width=\columnwidth]{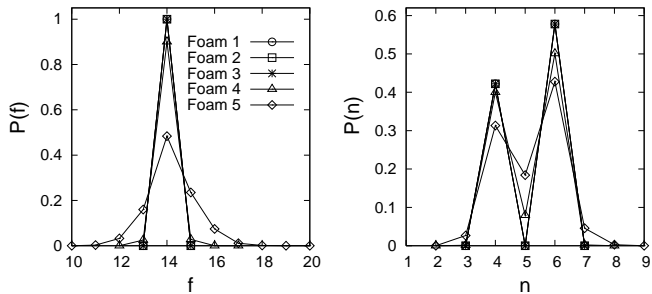}
\caption{Distribution of cells with $f$ faces, and distribution of
faces with $n$ edges, for the disordered Voronoi foams 1-5.}
\label{v3}
\end{figure}

The distribution of $L$, $S$, $V$, $f$ and $n$ for our Voronoi
foams are depicted in Figs.\ \ref{v1}-\ref{v3}. Note that these
foams are constructed from an initial Kelvin structure, thus here
$L_{\text{Kelvin}}$, $S_{\text{Kelvin}}$, and $V_{\text{Kelvin}}$
serve as natural units of length, surface, and volume,
respectively. Characteristics of the Kelvin foam, our Voronoi
foams 1-5, real foams\ \cite{mon}, simulated foams\ \cite{kray1},
and the Poisson Voronoi tessellation\ \cite{okabe} are summarized
in Table\ \ref{t1}. We shortly discuss several points. The
standard deviations and variances in Table\ \ref{t1} and the
distributions in Figs.\ \ref{v1} and \ref{v2} clearly demonstrate
that the disorder in the Voronoi foams increases with increasing
displacement $h$ of the seed points. While Voronoi foam 1
($h/H_{\text {Kelvin}} = 0.02$) with its narrow distributions is
still close to the Kelvin foam, the widths of the distributions
grow with increasing $h$ and are quite broad in Voronoi foam 5
with $h/H_{\text {Kelvin}}=0.20$. This is especially visible in
the distributions for edge length $L$, cell surface $S$, and cell
volume $V$ in Figs.\ \ref{v1} and \ref{v2}. For $h/H_{\text
{Kelvin}} > 0.20$, the Voronoi foam approaches the Poisson Voronoi
tessellation. The unit cell of a Kelvin foam consists of eight
hexagonal and six square faces. This is especially visible in the
bimodal distribution for the number of faces with $n$ edges in
Fig.\ \ref{v3}, which occurs in all of our Voronoi foams 1-5. Note
that foam 4 additionally contains pentagons and in foam 5
furthermore triangles and heptagons appear. In real foams,
however, the distribution has its single maximum at $n=5$.

Our Voronoi foams constructed from an initial Kelvin structure,
are indeed distinct from real foams, although some of the
statistical characteristics of foams 4 and 5 are close to or agree
with the one of real foams. Nevertheless, as we have just
discussed, our Voronoi foams 1-5 are characterized by increasing
topological and geometrical disorder. In the following, we
investigate whether this disorder does have any influence on the
transport-mean-free path $l^{\ast}$.

\section{Diffusive Transport of Light}

We investigate the transport-mean-free path $l^{\ast}$ for two models.

\subsection{Model of Constant Intensity Reflectance} \label{rconst}

To explore the impact of disorder on $l^{\ast}$, we assume here
that liquid films have a constant reflectance $r$. We model single
photon paths in a Voronoi foam as a random walk with rules
motivated by ray optics, {\i.e.}, an incoming light beam is
reflected from a face with a probability $r$ or it traverses the
face with a probability $t=1-r$. This naturally leads to a
persistent random walk of the photons\ \cite{miriA}, where the
walker remembers its direction from the previous step\ \cite{kehr,
w1}. Persistent random walks are employed in biological problems\
\cite{furth}, turbulent diffusion \cite{tay}, polymers\
\cite{flory}, Landauer diffusion coefficient for a one-dimensional
solid\ \cite{55}, and in general transport mechanisms\ \cite{t0}.
More recent applications are reviewed in\ \cite{w2}.

Our computer program takes $10^4$ photons at an initial position,
and launches them in a direction specified by polar angles $\theta
$ and $\varphi $. Then it generates the trajectory of each photon
following a standard Monte Carlo procedure and evaluates the
statistics of the photon cloud at times $\tau \in [10000,10200
,...,15800]$ (in units of $ \langle L \rangle /c $, where $c$
denotes the velocity of light). As detailed in Ref.\ \cite{miriA},
we determine the diffusion constant $D$ from the temporal
evolution of the average mean-square displacement of the photons:
$\langle \mathbf{r}^2\rangle = 6 D t$. Then the
transport-mean-free path follows from
\begin{equation}
l^{\ast} = 3 D/c .\label{lkj}
\end{equation}

\begin{figure}
\includegraphics[width=0.75\columnwidth]{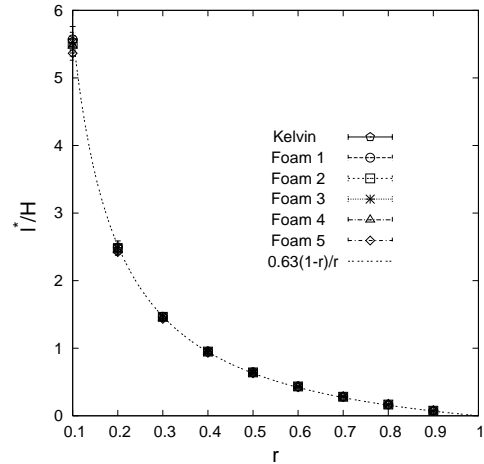}
\caption{ The transport-mean-free path $l^{\ast}$ (in units of the
average cell diameter $H$) as a function of intensity reflectance
$r$, for the Kelvin and the disordered Voronoi foams 1-5. The
Monte Carlo simulation results and the fit $l^{\ast}(r)/H =
0.63(1-r)/r$ are denoted, respectively, by points and line. }
\label{curveDr}
\end{figure}

\begin{figure}
\includegraphics[width=0.75\columnwidth]{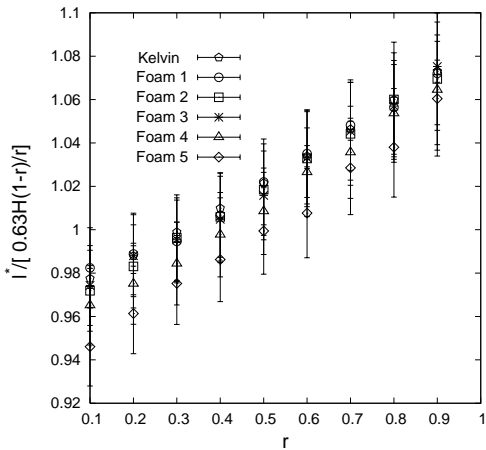}
\caption{The transport-mean-free path $l^{\ast}$ plotted relative
to $0.63 H (1-r)/r$ as a function of intensity reflectance $r$.
We find that $l^{\ast}_{\text{Voronoi}}(r)= 0.63 H ({1-r})/{r} ~(1-b_1  + b_2
r)$ with $0<b_{1}<0.1$ and $b_{2} \approx 0.1$.
 } \label{curveDrscaled}
\end{figure}

For angles $ \theta\in [3^{\circ}, 20^{\circ}, ..., 71^{\circ} ]$
and $\varphi \in[3^{\circ}, 20^{\circ}, ..., 156^{\circ}] $, the
simulation is repeated for each intensity reflectance $r \in [0.1,
0.2,...,0.9]$. As a reasonable result, no dependence on the
starting point and the starting direction is observed. In Fig.\
\ref{curveDr} we plot the transport-mean-free path $l^{\ast}$ in
units of the average cell diameter $H$ as a function of $r$, where
$H$ is defined via $H = [6 \langle V \rangle /\pi]^{1/3}$. The
line is a fit to $ 0.63  (1-r)/r$. In Fig.\ \ref{curveDr} the
transport-mean-free paths $l^{\ast}$ for the regular Kelvin foam
and the disordered Voronoi foams 1-5 are {not} distinguishable
from each other. To increase the resolution, the rescaled
transport-mean-free path $ l^{\ast}(r)/[0.63 H (1-r)/r]$ versus
$r$ is plotted in Fig.\ \ref{curveDrscaled}. The errorbars reflect
the the standard deviations when we average over all
transport-mean-free paths $ l^{\ast}(\theta,\varphi)$ for
different starting positions and angles. From Fig.\
\ref{curveDrscaled} we find that our numerical results agree well
with the relation $l^{\ast}_{\text{Voronoi}}(r)= 0.63 H
\frac{1-r}{r}(1-b_1  + b_2 r)$ mentioned already in Eq.
(\ref{emp}) in the introduction. Both constants $b_1$ and $b_2$
show a slight dependence on disorder in the Voronoi foam. The
results here already demonstrate that disorder in the Voronoi
foams does not change $l^{\ast}$ significantly.

\subsection{Model of Thin-Film Intensity Reflectance} \label{FreFre}

In the second, more realistic model, we use the intensity
reflectance of thin films. Again, we study the persistent random
walk of photons to probe the influence of disorder in the Voronoi
foams on the transport-mean-free path.

For a plane wave with wave vector $k \hat{\bm{k}} $ incident from
the air onto a liquid film with normal vector $\hat{\bm{n}} $, the
incident electric field with a general state of polarization is
$\bm{E}_{\text{incident}}= E_1 e^{\imath \phi_1} \hat{\bm{e}}_{1}
+ E_2 e^{\imath \phi_2} \hat{\bm{e}}_{2}+ E_3 e^{\imath \phi_3}
\hat{\bm{e}}_{3}$, where $E_m $ and $\phi_m$ are respectively the
magnitude and phase of the field component along the unit vector
$\hat{\bm{e}}_{m}$ ($m=1,2,3$). Taking into account all possible
multiple refraction paths in the film\ \cite{reitz}, the electric
field vectors of the transmitted and reflected waves become
\begin{eqnarray}
 \bm{E}_{\text{transmitted}} &=& (t_s-t_p )(\bm{E}_{\text{incident}} \cdot
\hat{\bm{b}}) \hat{\bm{b}} +t_p \bm{E}_{\text{incident}}\nonumber \\
\bm{E}_{\text{reflected}} &=& (r_s+r_p)(\bm{E}_{\text{incident}}
\cdot \hat{\bm{b}}) \hat{\bm{b}}
-r_p \bm{E}_{\text{incident}} \nonumber \\
& &  + 2 r_p (\bm{E}_{\text{incident}} \cdot \hat{\bm{n}}) \hat{\bm{n}} ,
\label{elecf}
\end{eqnarray}
from which one calculates the intensity reflectance
$r(i) = |\bm{E}_{\text{reflected}}|^{2}/|\bm{E}_{\text{incident}}|^{2}$ as
a function of the incident angle $i$ of the plane wave:
\begin{eqnarray}
r(i) = |r_p|^2 +\frac{ |\bm{E}_{\text{incident}} \cdot
\hat{\bm{b}}|^2}{ | \bm{E}_{\text{incident}}|^2} (|r_s|^2-|r_p|^2).
\label{elecr}
\end{eqnarray}
Here $|~~ |$ denotes the magnitude of a complex number, and
$\hat{\bm{b}} =\hat{\bm{n}} \times \hat{\bm{k}}$. Details of our
approach with the explicit formulas for the coefficients $r_p$,
$t_p$, $r_s$ and $t_s$ are given in Ref.\ \cite{miriD}. We
implemented the reflectance $r(i)$ of Eq.\ (\ref{elecr}) in our
Monte Carlo simulations. Again, we used the launching directions
for the photons as mentioned in Sec.\ \ref{rconst}, but evaluated
the statistics of the photon cloud at times $\tau \in [30000,
30200,...,35800]$ (in units of $\langle L \rangle /c$). Note that
the thin-film reflectance is small except near the grazing
incidence, thus long simulation times are required to achieve the
accuracy reported in Sec.\ \ref{rconst}.

The reflectance $r$ crucially depends on the ratio $d/\lambda$ of
film thickness $d$ and wavelength $\lambda$ of light, the
refractive index $n_0$ of the film, and the incident angle $i$.
Note that even in films as thin as the common black film, $r$
increases to $1$ close to {grazing} incidence ($ i \rightarrow
90^{\circ} $). This feature basically explains why films
significantly contribute to the diffusion of light in aqueous
foams\ \cite{miriB,miriD}.

\begin{figure}
\includegraphics[width=0.85\columnwidth]{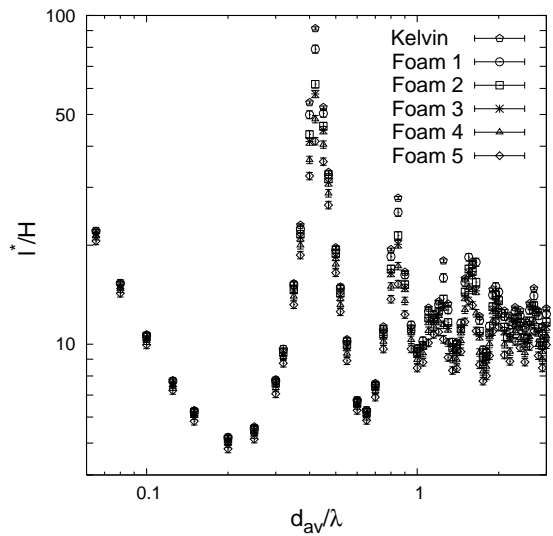}
\caption{The transport-mean-free path $l^{\ast}$ (in units of the
average cell diameter $H$) as a function of $d_{av}/\lambda$ for
the Kelvin foam and disordered Voronoi foams 1-5. The refractive
index $n_0$ of the liquid phase is $1.34$.} \label{ggg1}
\end{figure}

\begin{figure}
\includegraphics[width=0.7\columnwidth]{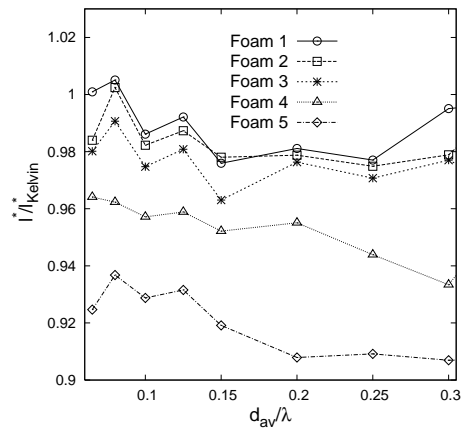}
\caption{The transport-mean-free path $l^{\ast}$  of Voronoi foams
1-5 plotted relative to $l^{\ast}_{\text{Kelvin}}$ as a function
of $d_{av}/\lambda$. For all these foams, $d_w=0$.} \label{com}
\end{figure}

\begin{figure}
\includegraphics[width=0.85\columnwidth]{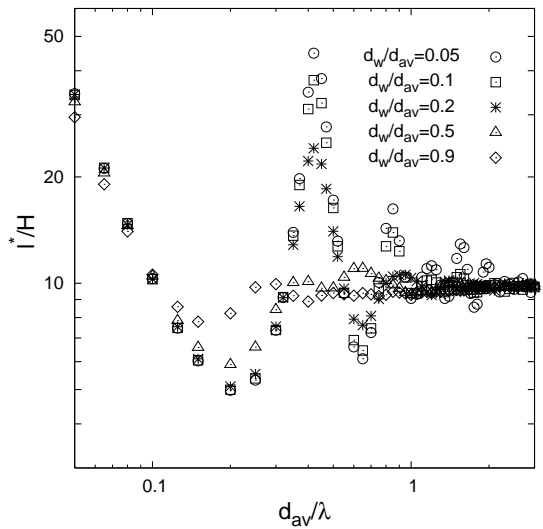}
\caption{The transport-mean-free path $l^{\ast}$ (in units of the
average cell diameter $H$) as a function of $d_{av}/\lambda $ for
Voronoi foam 4 with various thickness distribution. Note, for
$d_w/d_{av}=0.5$ and $0.9$, the pronounced maximum disappears.}
\label{Fresneldisor}
\end{figure}

First, we assume that all films of the foam have the same
thickness $d_{av}$. In Fig.\ \ref{ggg1} we plot $l^{\ast}$ as a
function of $d_{av}/\lambda$ for the Kelvin foam and our
disordered Voronoi foams 1-5. For $d_{av}/\lambda <0.2 $, the
transport-mean-free path monotonically decreases as $d_{av}$
increases. There is a pronounced minimum around $d_{av}/\lambda
=0.2$. Between $d_{av}/\lambda =0.2$ and $3$ oscillations exist
that are due to oscillations in the reflectance $r$ of a thin
film. The oscillations are {reduced} for more disordered foams but
do not disappear. The transport-mean-free path $l^{\ast}$ exhibits
two strong maxima around $d_{av}/\lambda =0.43$ and $0.84$. A
closer inspection of Fig.\ \ref{ggg1} shows that the heights of
the pronounced maxima at $d_{av}/\lambda =0.43$ and $0.84$
decrease as the disorder increases.
To explain this behavior, we note that in the Kelvin structure,
long straight photon paths can cross parallel faces. These long
photon paths occur for the parallel polarization state since below
the Brewster angle the reflectance is small and especially for the
most pronounced maximum at $d_{av}/\lambda = 0.43$ close to zero\
\cite{miriD}. With increasing disorder in the Voronoi foams these
parallel faces and, therefore, long straight photon paths,
disappear. In Fig.\ \ref{ggg1} the transport-mean-free paths
$l^{\ast}$ for the regular Kelvin foam and the disordered Voronoi
foams 1-5 are not distinguishable from each other in the region of
$d_{av}/\lambda < 0.3$. To increase the resolution, the rescaled
transport-mean-free path $l^{\ast}/l^{\ast}_{\text{Kelvin}}$ of
Voronoi foams 1-5 is plotted for this region in Fig.\ \ref{com}.
Clearly, the transport-mean-free paths differ from each other by
at most 10 \%.

Second, in our model we introduce some additional randomness in
the thickness $d$ of the films assuming that it is uniformly
distributed in $[d_{av}-d_w, d_{av}+d_w]$, where $d_{av}$ denotes
the average thickness and $d_w$ the width of the distribution.
In Fig.\ \ref{Fresneldisor} we plot $l^{\ast}$ as a function of
$d_{av}/\lambda $ for Voronoi foam 4 with various thickness
distributions. Other disordered Voronoi foams show the same
behavior. Obviously, disorder in $d$ decreases the oscillations to
an approximately constant transport-mean-free path for
$d_{av}/\lambda >1$. The two strong maxima are reduced noticeably
and ultimately disappear for strong disorder in the film
thickness, as illustrated by the data for $d_w/d_{av}=0.5$ and
$0.9$. However, we still observe that moderate disorder in the
thickness ($d_w/ d_{av} \leqslant 0.2$) does not affect $l^{\ast}$
and its monotonic behavior for $d_{av}/\lambda <0.2$.

\section{Discussion, Conclusions, and Outlook}\label{dis}

To understand the role of liquid films for light transport in dry
foams, we started with the two-dimensional honeycomb\ \cite{miriA}
and Voronoi structures\ \cite{miriB}. Then we focused on the
Kelvin foam\ \cite{miriD} to explore the effect of space
dimension. The Kelvin foam is periodic in space, while real foams
are disordered. To investigate the influence of topological and
geometrical disorder on diffusive photon transport, we have
utilized in this paper the three-dimensional Voronoi
tessellations.

We have studied the photon's persistent random walk in a
three-dimensional Voronoi structure based on rules motivated by
ray optics. In a first model, we used a constant intensity
reflectance $r$. The interesting result is that the
transport-mean-free path for the honeycomb structure, the
two-dimensional Voronoi foam, the three-dimensional Kelvin
structure, and the three-dimensional Voronoi foam are all
determined by the same factor $(1-r)/r$ for constant intensity
reflectance $r$ in spite of the differences in {\it dimension} and
{\it structure}. For the two-dimensional Voronoi foams, we even
see the same behavior as in Eq.\ (\ref{emp}) but with the
prefactor $0.63$ replaced by 0.55. Finally, the results here
already demonstrate that disorder in the Voronoi foams does not
change $l^{\ast}$ significantly [see Figs.\ \ref{curveDr} and\
\ref{curveDrscaled}]. This confirms our speculation in Refs.\
\cite{miriA,miriB,miriC,miriD} that neither the dimension of space
nor disorder have a strong influence on the magnitude of the
transport-mean-free path. Note that the factor $(1-r)/r$ expresses
the fact that independent of the dimension of space and the
different shapes of the cells in a foam, photon transport is
ballistic for $r=0$, and that photons stay confined to the initial
cell for $r=1$.

In a second, more realistic model, we used the intensity
reflectance of a thin film, with its significant dependence on
film thickness $d_{av}$ and angle of incidence $i$. Close to
grazing incidence ($ i \rightarrow 90^{\circ} $), the reflectance
always sharply increases to one. Thus, a thin film, even as thin
as the common black film, randomizes the photon direction and
contributes to $l^{\ast}$. Based on extensive Monte-Carlo
simulations, we paid special attention to the behavior of
$l^{\ast}$ at small $d_{av}/\lambda$: in real foams $d_{av}<100\
\mathrm{nm}$\ \cite{Vera2002} and, combined with the visible
portion of the spectrum ($ 450 \ \mathrm{nm} < \lambda < 750 \
\mathrm{nm}$), one arrives at $d_{av}/\lambda <0.2$ as the
relevant region in explaining the measurements of the
transport-mean-free\ path in Ref.\ \cite{durApp}. Quite
interesting, we found that the monotonic behavior of $l^{\ast}$
for $d_{av}/\lambda <0.2$ does {not} show a significant dependence
on the detailed geometrical structure of our Voronoi foams and on
moderate disorder in the film thickness [see Figs.\
\ref{ggg1}-\ref{Fresneldisor}].

Real foams obey Plateau's laws. They state that the following: i)
each film has a constant mean curvature; ii) three films meet at
angles of $120^{\circ}$ at each Pleateau border; and iii) these
borders meet in fours at the tetrahedral angle [$\arccos(-1/3)=
109.47^{\circ}$] to form a vertex. A Kelvin foam with flat faces
of its cells approximately obeys the second and third law;
however, a disordered Voronoi foam does not. We found that the
transport-mean-free path of the Kelvin and the Voronoi structures
deviate from each other by at most 10\% in the experimentally
relevant region for $d_{av}/\lambda$. This is an indication that
fulfilling Plateau's second and third laws is of minor importance
when one tries to understand the role of liquid films for
diffusive transport of light in foams. Even the slight curvature
of the films might not be very important since in the ray optics
approach it just introduces some additional randomness in the
relevant surface normal which varies across the film. Future
investigations on realistic foam structures, e.g., constructed by
the Surface Evolver will clarify all these aspects.

Vera, Saint-Jalmes, and Durian observed the empirical law stated
in Eq.\ (\ref{emp}) for  $0.008<\varepsilon<0.3$. Experimental
values for $l^{\ast}/H$ increase from 2 to 20 for decreasing
$\varepsilon$, whereas we determine a range of $l^{\ast}/H$
between $5$ (for $d_{av}/\lambda =0.2$) and 25 (for
$d_{av}/\lambda =0.06$, i.e., for a common black film) as
illustrated in Fig.\ \ref{ggg1}. That means our theoretical values
for $l^{\ast}$ fall in the same range as the experimental values
for the driest foams in Ref.\ \cite{durApp}. This shows that
liquid films are important for the understanding of photon
diffusion in dry foams otherwise we should have obtained much
larger transport-mean-free paths. 
However, to compare our theoretical findings to experiments, 
we should know the contribution of the Plateau borders to $l^{\ast}$. 
After all, even in a rather dry foam with a liquid fraction
of 1 \%, most of the liquid resides in these borders
and their junctions. Furthermore, a relation between the
film thickness $d_{av}$ and the liquid volume fraction $\varepsilon$
is needed.
To the best of our knowledge, systematic
measurements of the film thickness in dry foams have not been
reported, neither do theoretical models exist. In the case of
dense oil-in-water emulsions, the dependence of the film thickness
on the oil-volume fraction has been modeled by Buzza and Cates\
\cite{bu}. However their results cannot be applied directly to
foams.


In this paper, we focused on the contribution of liquid films to
the transport-mean-free path of photon diffusion in {dry} foams.
To achieve a complete understanding of the subject, there is still 
much to do. Of immediate interest are modeling of the film thickness 
variation with the liquid volume fraction and following the random walk 
of photons by taking into account {\it both} their refraction at 
thin {\it films} and their scattering from {\it Plateau borders}.
To the best of our knowledge, the scattering matrix for a Plateau
border is not available. However, ray optics reveals intriguing
optical properties of the Plateau borders\ \cite{sun1, sun2}.
Using the hybrid lattice gas model for two-dimensional foams, Sun
and Hutzler\ \cite{sun1} emphasize that the complex geometry of
the Plateau borders affects the optical properties of foams.
However, Plateau borders of a two-dimensional foam do not form a 
network and channeling of the photons, as observed in 
Ref.\ \cite{Gittings2004}, is impossible. 
Thus an extension of Ref.\ \cite{sun1} to three-dimensional foams
would be interesting.


\begin{acknowledgments}
MF.M. and Z.S. thank Iran's Ministry of Science, Research and
Technology for support of the parallel computing facilities at
IASBS under Grant No. 1026B (503495), Iranian Telecommunication
Research Center (ITRC) for financial support, and Farhad Abdi for
technical assistance. H.S. appreciates discussions with Regine von
Klitzing.
\end{acknowledgments}

\end{document}